# Continuous phase control of nonlinear polarizability in harmonic generations

Guixin Li[1,2], Shumei Chen[1,2], Nitipat Pholchai[3,4], Polis Wing Han Wong[5], Edwin Yue Bun Pun[5], KokWai Cheah[2]*, Thomas Zentgraf[3]*, and Shuang Zhang[1]*

[1]School of Physics & Astronomy, University of Birmingham, Birmingham, B15 2TT, UK

[2]Department of Physics, Hong Kong Baptist University, Kowloon Tong, Hong Kong

[3]Department of Physics, University of Paderborn, Warburger Straße 100, D-33098 Paderborn, Germany

[4]Faculty of Applied Science, King Mongkut's University of Technology North Bangkok, Bangkok 10800, Thailand

[5]Department of Electronic Engineering, City University of Hong Kong, 83 Tat Chee Ave, Hong Kong

*Correspondence to: kwcheah@hkbu.edu.hk; thomas.zentgraf@uni-paderborn.de; s.zhang@bham.ac.uk;

**Abstract**: We prescribe a novel approach for continuously tailoring the local phase of the nonlinear polarizability which can lead to an arbitrary phase profile for harmonic generations. The introduced phase of the nonlinear polarizability is inherently a geometric Berry phase arising from the spin rotation coupling of light in the nonlinear regime. This approach provides new routes for controlling the optical nonlinear processes.

The capability of locally engineering the nonlinear optical properties of media is crucial in nonlinear optics. The most well-known example is quasi-phase matching in second order nonlinear processes. The quasi-phase matching leads to efficient frequency conversion compared to a homogeneous nonlinear medium by providing the extra momentum to compensate the phase mismatch between the fundamental and harmonic waves [1, 2]. Quasi-phase matching has been used for various applications, ranging from beam and pulse shaping [3], multiple harmonic generation [4], nonlinear Airy beam generation [5], to generation of entangled photons [6]. Poling is the most widely employed technique for achieving quasi-phase matching [7]. By periodically reversing the crystalline orientation of ferroelectric materials, the sign of the $\chi^{(2)}$ nonlinear coefficient can be spatially modulated along the propagation direction. However, poling only leads to a binary state for the nonlinear material polarization, which is equivalent to a discrete phase step of $\pi$ of the nonlinear polarization. Here we propose a nonlinear metamaterial with homogeneous linear optical properties but continuously controllable phase of the nonlinear polarizability. Such a nonlinear metamaterial would enable exact phase matching conditions for nonlinear optical processes, hence surpassing the efficiency of the widely utilized quasi-phase matching scheme in which only the sign of the nonlinear polarizability can be manipulated. More importantly, it may remove additional undesired nonlinear processes which are unavoidable in a periodically poled nonlinear system [2]. Further, it is expected that the nonlinear metamaterial may enable complete control of the propagation direction of the nonlinear generated wave, and as constituent it may greatly enrich the functionalities of nonlinear photonic crystals.

The phase control over the nonlinear coefficient of the metamaterial is inspired by the concept of geometric Berry phase of light when it experiences a change in the polarization state [8]. Geometric phases play an important role in various branches of physics from condensed matter systems to optics. In optics, the geometric phase can be manifested as spin orbital coupling leading to optical spin selective behaviors such as spin Hall effect of light, and spin selective anomalous refraction [9]. Recently, the concept of geometric phase has been applied to the design of novel types of metasurfaces for controlling the phase of scattering at an interface [10-15]. These types of metasurfaces, which consist of plasmonic structures with subwavelength feature size (sometimes called "artificial atoms"), can be engineered to show rotation controlled local geometric phase shifts [12-14]. In the linear optical regime, this concept has been employed for flat lens imaging, generation of vortex beams, three dimensional holography, and optical spin orbital interaction, etc [12-15]. Here we will extend the concept of the Berry or geometric phase introduced by the light interaction with the metasurface to the nonlinear regime leading to a nonlinear material polarization with arbitrarily controllable phase profile.

We start by considering a single subwavelength plasmonic nanostructure embedded in an isotropic nonlinear medium (Fig. 1). We show that when excited by a circularly polarized fundamental beam, the phase of the nonlinear polarization of the artificial atom can be controlled geometrically by the orientation of the plasmonic structure through a spin rotation coupling. For incident light with the circular polarization state $\sigma$ propagating along +z direction, the electric field can be expressed as: $\vec{E}^\sigma = \tilde{E}_0(\vec{e}_x + i\sigma\vec{e}_y)/\sqrt{2}$, where $\sigma = \pm 1$ represents the state of left or right handed circular polarization, respectively. The plasmonic structure together with a nonlinear medium in close vicinity of the structure where the field is strongly enhanced, forms a local nonlinear dipole moment

$$\vec{p}_\theta^{n\omega} = \ddot{\alpha}_\theta(\vec{E}^\sigma)^n \qquad (1)$$

where $\alpha_\theta$ is the $n^{th}$ harmonic nonlinear polarizability tensor of the nanostructure with orientation angle of $\theta$. We employ a coordinate rotation to analyze the dependence of the nonlinear dipole moment on the orientation angle of the plasmonic structure. In the local coordinate of the plasmonic structure (referred to as local frame) as shown in Fig. 1(a) where the local coordinate (x, y) axes are rotated by an angle of $\theta$ with respect to the laboratory frame, the fundamental wave acquires a geometric phase due to the rotation spin coupling effect

$$\vec{E}_L^\sigma = \vec{E}^\sigma e^{i\sigma\theta} \qquad (2)$$

where the index '$L$' denotes the plasmonic structure's local coordinate frame. The $n^{th}$ harmonic nonlinear polarizability in the plasmonic structure's local frame is simply $\alpha_0$. Thus, the $n^{th}$ harmonic nonlinear dipole moment in the local frame is given by

$$\vec{p}_{\theta,L}^{n\omega} = \ddot{\alpha}_0(\vec{E}_L^\sigma)^n = \ddot{\alpha}_0(\vec{E}^\sigma)^n e^{in\sigma\theta} \qquad (3)$$

The nonlinear dipole moment can be decomposed into two in-plane rotating dipoles as

$$\vec{p}_{\theta,L}^{n\omega} = \vec{p}_{\theta,L,\sigma}^{n\omega} + \vec{p}_{\theta,L,-\sigma}^{n\omega} \text{ with } \vec{p}_{\theta,L,\sigma}^{n\omega}, \vec{p}_{\theta,L,-\sigma}^{n\omega} \propto e^{in\sigma\theta} \quad (4)$$

After transforming back to the laboratory frame the two rotating dipole moments are given by

$$\vec{p}_{\theta,\sigma}^{n\omega} = \vec{p}_{\theta,L,\sigma}^{n\omega} e^{-i\sigma\theta} \propto e^{(n-1)i\sigma\theta}$$
$$\vec{p}_{\theta,-\sigma}^{n\omega} = \vec{p}_{\theta,L,-\sigma}^{n\omega} e^{i\sigma\theta} \propto e^{(n+1)i\sigma\theta} \quad (5)$$

Thus, a nonlinear geometric Berry phase with a phase variation of $(n-1)\sigma\theta$ or $(n+1)\sigma\theta$ is introduced to the $n^{th}$ harmonic generation for the same or the opposite circular polarization to that of the fundamental wave, respectively. According to the selection rules for harmonic generation of circular polarized fundamental waves, a plasmonic nanostructure with m-fold rotational symmetry only allows harmonic orders of $n = lm \pm 1$, where $l$ is an integer, and the '+' and '-' sign corresponds to harmonic generation of the same and opposite circular polarization, respectively [16-18]. Hence for a plasmonic nanorod structure with two-fold rotational symmetry (C2), THG signals with both same and opposite circular polarizations as the fundamental wave can be generated. According to Eq. 5, they have a spin dependent phase of 2σθ and 4σθ, respectively. On the other hand, a plasmonic nanostructure with 4-fold rotational symmetry (C4) does not allow a THG process for the same polarization state as the incident polarization. Hence, only a single THG signal, that of the opposite circular polarization, is generated with a geometrical phase of 4σθ. Importantly, the circular polarization of the fundamental wave remains the same after transmitting through a metamaterial consisting of such C4 nanostructures of arbitrary orientations. Thus, by assembling the C4 nanostructures with spatially varying orientations in a 3D or 2D lattice, a nonlinear metamaterial or metasurface can be formed which show homogeneous linear properties, but locally a well-defined nonlinear polarization distribution for a circularly polarized fundamental wave, as schematically illustrated in Fig. 1(b) and Fig. 1(c).

To conclude, we showed that a Berry phase can be utilized to continuously tune the phase of the nonlinear polarization in a metamaterial or metasurface consisting of nanocrosses with spatially varying orientations. The metamaterials with custom defined local nonlinear susceptibilities may introduce new freedoms in designing nonlinear materials for satisfying perfect phase matching condition, nonlinear photonic crystals with arbitrary nonlinear phase profiles, and manipulation of nonlinear signals in integrated photonic circuits or in the free space.

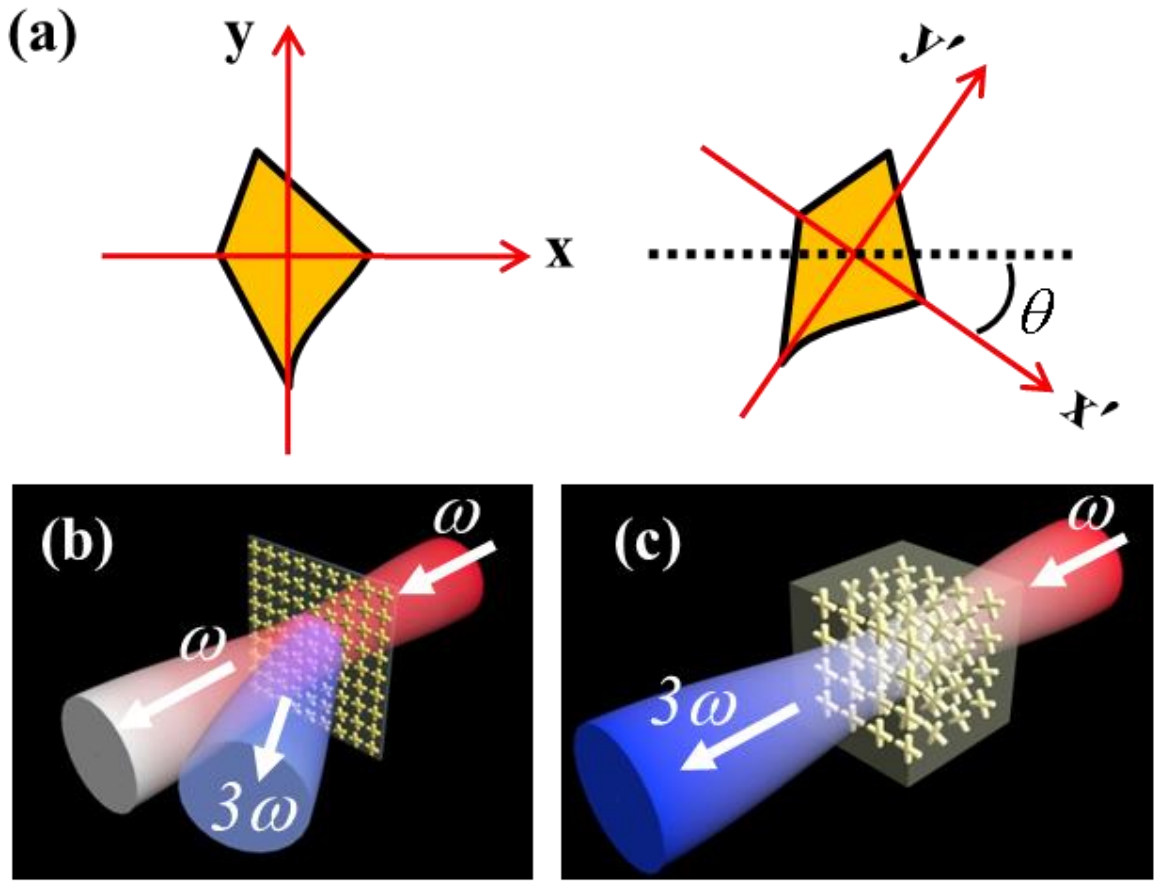

**Fig. 1. Illustration of geometrical phase controlled nonlinear metamaterials.** (a) Rotating an nanostructure by angle θ with respect to the laboratory frame will introduce a geometrical phase: *θ*; each nanostructure introduces nonlinear geometric Berry phase with a phase variation of (n-1)σθ or (n+1)σθ to the nth harmonic generation for the same or the opposite circular polarization to that of the fundamental wave, respectively. (b) A metasurface with an in-plane phase gradient of nonlinear polarizability for controlling the direction of the third harmonic generation signal (*n*=3). Due to the C4 rotation symmetry of the individual nanostructures, the fundamental beam transmission is not affected by the rotation gradient. (c) A 3D nonlinear metamaterial with a continuous phase modulation of nonlinear polarizability along the propagation direction of fundamental wave for obtaining perfect phase-matching in the third-harmonic generation process.